\documentstyle[11pt,paspconf,psfig]{article}

\def\etal{et al.}
\def\hii{H{\sc ii}}

\def\msun{M$_{\odot}$}
\def\zsun{Z$_{\odot}$}
\def\halpha{\ifmmode {\rm H{\alpha}} \else $\rm H{\alpha}$\fi}
\def\hbeta{\ifmmode {\rm H{\beta}} \else $\rm H{\beta}$\fi}
\def\heii{\ifmmode {{\rm He{\sc ii}} \lambda 4686} \else {He\,{\sc ii} $\lambda$4686}\fi}

\def\oiii{[O\,{\sc iii}]}

\def\niii{N\,{\sc iii} $\lambda$4640}

\def\ciii{C\,{\sc iii} $\lambda$4650}

\def\civ{C\,{\sc iv} $\lambda$5808}

\begin{document}

\title{Wolf-Rayet galaxies as probes of young stellar systems
\footnotemark[1]} 

\author{Daniel Schaerer} 
\affil{Observatoire Midi-Pyr\'en\'ees, 
       14, Av. E. Belin, F-31400, Toulouse, France
      (schaerer@obs-mip.fr)}

\footnotetext[1]{To appear in "SpectroPhotometric Dating of Stars and Galaxies", 
eds. I. Hubeny, S. Heap, and R. Cornett, ASP Conf. Series, in press}

\begin{abstract}
Wolf-Rayet (WR) galaxies provide detailed information on massive star populations
in starbursts and thereby represent ideal objects to determine accurate ages for 
young systems, to measure the burst duration, and to probe the upper end of the IMF.
WR galaxies play also a particular role in a variety of studies on
star formation in Seyfert2/LINERS, dust production and local chemical enrichment,
temperature fluctuations in \hii\ regions, gaz outflows and X-ray emission etc.

Different age indicators for young starbursts and WR galaxies are discussed.
We summarise recent work on the burst properties of WR galaxies, massive star 
populations at different metallicities, and the use of WR galaxies as benchmarks
for multi-wavelength models of starbursts and photoionisation models including 
IR fine structure lines.

\end{abstract}

\section{Introduction}
Among emission line galaxies, the so-called Wolf-Rayet (WR) galaxies are defined
by the presence of one or several {\em broad stellar emission lines} in the optical
(often called ``WR bump'') attributed to WR stars (Conti 1991, Schaerer \etal\ 1999b).
WR galaxies (or more accurately and independently of size, regions 
hosting detectable numbers of WR stars) are found among a large variety of objects 
including BCD, massive
spirals, IRAS galaxies, Seyfert 2, and LINERs (Schaerer \etal\ 1999b).

Together with the UV P-Cygni lines (mostly Si~{\sc iv}, C~{\sc iv}, but also O~{\sc vi};
all originating in O stars)
frequently observed in starbursts (e.g.\ Leitherer 1997, Gonz\'alez Delgado \etal\ 1998),
the WR lines are the most direct evidence of massive stars in these objects.
Other lines, like e.g.\ H and He optical absorption lines due to O stars or 
UV photospheric metal lines from B stars, are generally weaker and/or more 
difficult to observe (cf.\ Gonz\'alez Delgado \etal\ 1999, de Mello \etal\ 1999).
All the above lines can be used to determine quantitatively the massive star populations
(Wolf-Rayet, O, B) in the objects of interest. 
Compared to objects showing any of these spectral features, the 
main advantages of the more rare galaxies showing WR lines are the following:
WR stars being the descendents of the most massive stars and a short-lived phase
($M_{\rm ini} \ga 25$ \msun, $t_{\rm WR} \sim 10^{5-6}$ yr) their detection
provides an ideal age indicator for young systems ($t \la 10$ Myr), a good measure of 
the burst duration, and the best direct probe of the upper end of the IMF.

Given these basic considerations, WR galaxies are ideally used for studies
on age datation of young systems, determinations of fundamental burst
properties (IMF, SFR, etc.). They also allow to obtain constraints on stellar evolution
models, e.g.\ at very low metallicities and serve as benchmarks for the modeling
of UV-IR emission from starbursts. 
Results from such work is summarised in the present review.

In addition to these subjects,
WR star populations play a particular role in a variety of other studies on:
\begin{itemize}
\item the importance of star formation in Seyfert 2 and LINERS (Heckman \etal\ 1997)
\item dust production by WR stars (Williams 1995, Dwek 1998)
\item local chemical enrichment by WR stars (Walsh \& Roy 1987, Kobulnicky \& Skillman
  1996) and their possible influence on the primordial He abundance determination
  (Pagel \etal\ 1992, Esteban \& Peimbert 1995)
\item temperature fluctuations in \hii\ regions
  (Gonz\'alez-Delgado \etal\ 1994, P\'erez 1996, Luridiana \etal\ 1999)
\item the origin of optical--IR high excitation lines (Garnett \etal\ 1991, Schaerer
  1996, Schaerer \& Stasi\'nska 1999)
\item gaz outflows and X-ray emission from starbursts (e.g.\ Stevens \& Strickland 1998)
\end{itemize}
These issued will not, or only briefly, be discussed here.
The reader is referred to the above selected references for more information.
An overview of research on the WR phenomenon in stars and  galaxies is found
in the proceedings of the recent IAU Symposium 193 (van der Hucht \etal\ 1998).

\section{Datation of young objects}
A general review of the datation of young stars and stellar systems is presented by
Leitherer (these proceedings). Here we focus on WR galaxies and the youngest
objects.

{\bf Age from the WR features:} 
The presence of WR stars indicates recent massive star formation. 
Stellar models (single stars) predict WR stars between ages of $\sim$ 2 to 8 Myr, 
the latter limit strongly depending on metallicity (cf.\ Maeder \& Meynet 1994).
The effect of rotation leading to additional mixing processes can also
modify the lifetimes/ages of WR stars (cf.\ Meynet, these proceedings).
Larger ages (up to $\sim$ 10-30 Myr) can be obtained if WR stars form e.g.\ through 
Roche lobe overflow in massive close binary systems and possibly even from secondary 
mass gainer stars in such systems (see e.g. Vanbeveren \etal\ 1998, Mas-Hesse \& Cervi\~no
1998).
The exact binary scenarios and their frequency are not well established yet and fundamental
tests e.g.\ on clusters remain to be done.
Assuming that the single star channel dominates the WR production at most metallicities
(see e.g.\ Maeder \& Meynet 1994) we thus conclude that the presence of the stellar features
in WR galaxies indicates ages of 2-8 Myr for the bulk of this population.
Including information on the metallicity, more refined age estimates can be made using
evolutionary synthesis models (e.g.\ Schaerer \& Vacca 1998, hereafter SV98).

%
%

\begin{figure}[ht]
\centerline{
\psfig{figure=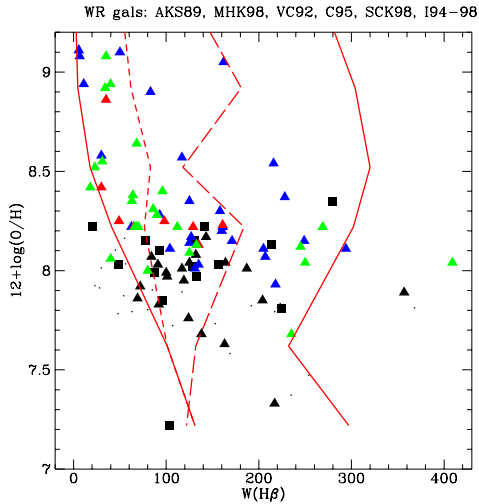,width=7cm}}
\caption{Position of WR galaxies in metallicity (O/H) vs.\ \hbeta\ equivalent
width (triangles, squares). Different greytones/colours indicate different samples.
Solid lines denote the beginning/end of the WR-rich phase predicted by the
SV98 models. Dashed lines indicate age steps of 1 Myr during the WR phase.
}
\label{fig_combine}
\end{figure}

{\bf The $W(\hbeta)$ age indicator:} 
The \hbeta\ equivalent width provides another potential age indicator for young
regions ionized by massive stars (cf.\ Copetti \etal\ 1986), which is commonly
predicted by evolutionary synthesis models.
In an instantaneous burst $W(\hbeta)$ decreases rapidly with time.
For obvious reasons the  $W(\hbeta)$--age relation is sensitive to the IMF slope and
its upper mass cut-off, metallicity, the exact star formation history, and depends
also on assumptions on the nebula (ionization boundedness etc.)
Furthermore the measured $W(\hbeta)$ can be reduced due to an underlying stellar population.
Although this indicator is well known, even for \hii\ regions no systematic comparisons 
have been undertaken to compare ages derived from $W(\hbeta)$ to those derived from their
stellar content.
However, it is well known that regions with very 
large \hbeta\ equivalent widths such as predicted for populations of 
ages $\sim$ 0-1 Myr are not observed. The origin of this 
discrepancy is not understood yet.

A comparison of the observed and predicted $W(\hbeta)$ for a large sample of WR galaxies
is shown in Fig.\ \ref{fig_combine} (see Schaerer 1998 for more details).
Note that the bulk of the objects shown here represent star forming regions
(e.g.\ giant \hii\ regions etc.) in BCDs or spirals.
The solid lines show the beginning and end of the WR phase for instantaneous
burst models with a Salpeter IMF and $M_{\rm up}=$ 120 \msun.
Fig.\ \ref{fig_combine} shows that essentially all WR galaxies lie within the predicted
W(\hbeta) range; in the most populated metallicity range (12+$\log$(O/H)
$\sim$ 7.8 -- 8.5) the observations also fully populate this domain.
It can thus be concluded that on average the 
{\em age and duration of the WR-rich phase predicted by the SV98 models 
for instantaneous bursts agree quite well with the observations.}
Despite the uncertainties mentioned above, the use of the $W(\hbeta)$ age indicator
appears to work reasonably well for the known WR galaxies.

{\bf Comparison with photometric indicators:}
Devost (1999) has recently proposed the use of (B-H) vs. (H-K) color-color diagrams 
for the datation of young populations which allow to probe ages similar to those
of WR galaxies.
For unknown extinction his method allows the discrimination of regions younger and older 
than 4 Myr. With the availability of data on extinction, the time resolution is
significantly improved.
As shown by Devost (1999) for observations of Arp 299 the method yields a good agreement 
when compared with ages derived from the \halpha\ equivalent width.
This age indicator appears also to be confirmed by the spectroscopic detection of WR 
stars in knots B and C by Bill Vacca (see Schaerer \etal\ 1999b).
Conversely BHK photometry could thus quite efficiently be used for a selection of 
regions hosting WR stars.

\section{Stellar populations: burst properties and constraints on stellar models}
In this Section we will summarise the main results on massive star populations
in WR galaxies (see also Schaerer 1998 for a recent review).

{\bf Burst properties of WR galaxies:}
Two main conclusions could already be drawn from the early work of Arnault \etal\ (1989):
{\em 1)} To reproduce the observed WR and O star populations bursts of very short
duration (``instantaneous'') are required.
{\em 2)} the observed trend of increasing WRbump/\hbeta\ intensities with metallicity
is understood by the metallicity dependence of mass loss leading to an increased
number of WR stars at high Z.
These finding were largely confirmed by Vacca \& Conti (1992) and all later studies.

More detailed studies of the massive star content of WR galaxies have been
undertaken by Schaerer (1996), Schaerer \etal\ (1999a, hereafter SCK99), Huang \etal\ (1999),
Mas-Hesse \& Kunth (1999) and Guseva \etal\ (1999).
The main results from the study of SCK99 are [where available comparison with other 
results given in parenthesis]:
%
{\em 1)} Essentially instantaneous bursts ($\Delta t \la$ 2-4 Myr) are required 
to reproduce the observed WR line intensities. [Same as other authors.]
{\em 2)} The majority of the observations can be reproduced with a Salpeter IMF.
No clear case requiring a flatter IMF is found.
[Similar result as Mas-Hesse \& Kunth; cf.\ however Huang \etal\ 1999].
{\em 3)} The IMF in these regions must be populated up to large masses.
For the case of I Zw 18, at $Z \sim$ 1/50 \zsun, $M_{\rm up}$ extending
to 120-150 \msun\ is found (de Mello \etal\ 1998).
{\em 4)} The relative populations of WN and WC stars detected clearly favour
the high mass evolutionary models (cf. below).

The above results hold over a fairly wide range of metallicities (1/50 $\la$ Z/\zsun 
$\la$ 1; cf.\ Schaerer 1998).
The findings on the IMF slope and the upper mass cut-off are in 
agreement with several independent studies, e.g.\ stellar counts, UV line profile 
modeling of starbursts, photoionization models for \hii\ galaxies etc. (references in
Schaerer 1998).
Interestingly, however, there are several studies indicating the possibility
of a lower value for $M_{\rm up}$ at metallicities above solar
(e.g.\ ULIRG, HII regions: Bresolin \etal\ 1999).
A search for WR stars in metal rich regions will place direct 
constraints on the upper end of the IMF in such environments.

{\bf Stellar evolution models at different metallicities:}
The WR populations in starbursts are often found in clusters or super star clusters.
In this case, and provided the bulk of the population has been formed coevally,
the analysis of integrated spectra can thus provide powerful constraints on 
stellar evolution models. The case of mostly extra-galactic \hii\ regions is summarised
in the following.

SCK99 obtained high S/N spectra with sufficient resolution to resolve the main
components of the WR bump (\heii, \niii + \ciii, and others) and including the ``red
WR bump'' (\civ) allowing the first detailed analysis of both the WN and WC star population
in their objects. 
The distinction of WN and WC subtypes is of special interest since in particular their 
relative number is very sensitive the detailed evolutionary scenarios
(mass loss, mixing; cf.\ Maeder \& Meynet 1994, Meynet these proceedings, SV98).
The analysis of SCK99 and de Mello \etal\ on I Zw 18 (1/50 \zsun) shows that stellar evolution 
models with high mass loss are clearly favoured, both to reproduce the observed WR and O populations 
and especially the WC/WN ratio.
A similar study was undertaken more recently by Guseva \etal\ (1999) who analyse the WN, WC
and O star content in a larger sample of WR galaxies. As above, good agreement is found with 
the high mass loss stellar models for the majority of objects. 
Guseva \etal\ (1999) find a possible underestimate of the WR population at very low metallicities
($\la 1/10$ \zsun).

It is understood that part of the ``requirement'' for the high mass loss (cf.\ Schaerer 1998) may be 
compensated by additional 
mixing processes leading to a similar prolongation of the WR phase (cf.\ Meynet, these proceedings).
In any case both the well known stellar census in the Local Group (cf.\ Maeder \& Meynet 1994)
and new data from integrated populations place important constraints on the evolutionary
models. The new studies considerably extend the range of available metallicities to very low $Z$.

\section{WR galaxies as benchmarks for UV--IR modeling}
Since WR galaxies provide quite severe constraints on age and their massive star population
they offer a special opportunity and testbed for starburst models, such as
multi-wavelength evolutionary synthesis models, photoionisation models etc.
A brief summary of such analysis undertaken recently shall be presented here.

{\bf Multi-wavelength analysis:}
A multi-wavelength study of 17 blue compact galaxies (BCD), the majority of them
WR galaxies, was presented by Mas-Hesse \& Kunth (1999).
The results on the IMF slope and burst duration in the WR galaxies confirms 
those discussed above (e.g.\ SCK99).
From UV--optical observations obtained through essentially identical apertures Mas-Hesse \& Kunth
(1999) find some objects with a non-negligible contribution from an older population to the optical 
light. They also confirm earlier findings showing a smaller extinction derived from the 
stellar continuum compared to Balmer emission lines.
Finally, the general agreement between predicted and observed far-IR emission suggests only a 
negligible fraction of hidden star formation in their objects.
On the other hand,
multi-$\lambda$ observations of the prototypical starburst NGC 7714, also a known WR galaxy, 
reveal an important deficit of UV light compared the population observed in the IR (Goldader \etal\
1999; see also Leitherer these proceedings).

{\bf Photoionisation models:}
Tailored photoionisation models of WR galaxies (or more precisely extra-galactic \hii\ regions
hosting WR stars) have recently been constructed for NGC 2363 and I Zw 18 by Luridiana \etal\ (1999) 
and Stasi\'nska \& Schaerer (1999) respectively.
Both studies make use of state-of-the art evolutionary synthesis models including appropriate
WR atmosphere models for the description of the radiation field.
The presence of WR signatures and/or nebular \heii\ emission attributed to these stars provides
an important constraint in both studies.

Adopting the metallicity of NGC 2363 derived by standard methods, Luridiana \etal\ (1999)
show that is not possible to reproduce the main optical emission line ratios considering a wide
range of parameter space (IMF, age, and nebular parameters).
From their failure the authors suggest a larger metallicity, which could
be due to temperature fluctuations in the nebula. 

Spectroscopic data and HST \halpha\ images of I Zw 18 (NW region) providing information on its 
stellar content, nebular density distribution, and line emission were used by Stasi\'nska \& 
Schaerer (1999) to construct a detailed photoionisation model of this well known object.
The main results from their study is that even taking strong deviations from the adopted ionizing
radiation field and additional X-rays into account, photoionisation models underpredict 
\oiii\ $\lambda$4363/5007,
similarly to the case of NGC 2363 above, which indicates a missing energy source.
This finding is of importance for elemental abundance determinations and our understanding of \hii\
regions in general.
The origin of this discrepancy (shocks, conductive heating, ?) remains to be understood.

{\bf Modeling IR emission lines:}
To study the use of IR fine structure lines as potential indicators of star formation properties 
(IMF, $M_{\rm up}$, age, etc.; cf.\ Lutz \etal\ 1998, Colbert \etal\ 1999) in obscured systems 
we have recently constructed a combined evolutionary synthesis $+$ photoionisation model for the
WR galaxy NGC 5253 observed by ISO (Schaerer \& Stasi\'nska 1999, hereafter SS99).
Modeling of this armorphous galaxy hosting two main ionizing clusters with a relative well known
stellar content should provide an ideal test-ground before stepping to more complex systems 
like M82, ULIRG etc.

As shown in SS99 the ionization structure of H, He, and O as revealed by the 
optical and IR emission lines is well reproduced. The presence of WR stars also explains
the high excitation of this objects as indicated by the \heii\ and O~[{\sc iv}] $\lambda$25.9 $\mu$m 
emission.
Interestingly the IR line ratios of [Ne~{\sc iii}]/[Ne~{\sc ii}], [Ar~{\sc iii}]/[Ar~{\sc ii}], 
and [S~{\sc iv}]/[S~{\sc iii}] are predicted too strong for a single ionisation parameter fitting
the above constraints. This implies that the use of these line ratios as diagnostics for the massive 
star population would lead to inconsistent results.
The differences may be due to a more complex galaxy structure than assumed in the model, 
inadequacies in the ionizing spectrum, or erroneous atomic data (SS99).
Work is in progress to establish how general this type of discrepancy is and to understand its origin.
As this first test shows, caution needs to be applied when interpreting IR fine structure lines
in terms of star formation properties.


\begin{acknowledgements}
DS acknowledges a grant from the Swiss National Foundation of Scientific
Research and financial support from LOC.
\end{acknowledgements}


\begin{references}
\reference Arnault, P., Kunth, D., Schild, H. 1989, \aap, 224, 73 
\reference Bresolin F., Kennicutt R.C., Garnett D.R., 1999, \apj, 510, 104
\reference Colbert, J.W., 1999, \apj, 511, 721
\reference Conti, P.S., 1991, \apj, 377, 115
\reference Copetti, M.V.F., Pastoriza, M.G., Dottori, H.A., 1986, \aap, 156, 111
\reference de Mello, D.F., Leitherer, C., Heckman, T., 1999, \apj, submitted
\reference de Mello, D.F., Schaerer, D., Heldman, J., Leitherer, C., 1998, \apj, 507, 199
\reference Devost, D., 1999, AJ, in press (astro-ph/9904019)
\reference Dwek, E., 1998, \apj, 501, 643
\reference Esteban, C., Peimbert, M., 1995, \aap, 300, 78
\reference Garnett, D.R., et al., 1991, \apj, 373, 458
\reference Goldader, J.D., et al., 1999, \apj, in preparation
\reference Gonz\'{a}lez Delgado, R.M., et al., 1994, \apj, 437, 239
\reference Gonz\'{a}lez Delgado, R.M., et al., 1998, \apj, 495, 698
\reference Gonz\'{a}lez Delgado, R.M., Leitherer, C., Heckman, T., 1999, \apjs, in press
\reference Guseva, N.G., Izotov, Y.I., Thuan, T.X., 1999, \apj, submitted 
\reference Heckman, T., et al., 1997, \apj, 482, 114
\reference Huang J.H., \etal, 1999, ApJ, 513, 215
\reference van der Hucht, K., Koenigsberger, G., Eenens, P.R.J., (eds.), 1998
``Wolf-Rayet Phenomena in Massive Stars and Starburst Galaxies'',
IAU Symp. 193, ASP Conf. Series, in press
\reference Kobulnicky, H.A., Skillman, E.D., 1996, \apj, 471, 211
\reference Leitherer, C., 1997, in ``The Ultraviolet Universe at Low and High Redshift'',
  Eds. W.H. Waller et al., Woodbury: AIP, 119
\reference Luridiana, V., Peimbert, M., Leitherer, C., 1999, \apj, submitted
\reference Lutz, D., et al., 1998, in ``Star Formation with the Infrared Space Observatory'',
J.L. Yun, R. Liseau, Eds., ASP Conf. Series, 132, 89
\reference Maeder, A., Meynet, G., 1994, \aap, 287, 803
\reference Mas-Hesse, J.M., Cervi\~no, M., 1998, 
in ``Wolf-Rayet Phenomena in Massive Stars and Starburst 
Galaxies'', IAU Symp. 193, ASP Conf. Series, in press
\reference Mas-Hesse, J.M., Kunth, D., 1999, \aap, in press
\reference Pagel, B.E.J., Simonson, E.A., Terlevich R.J., Edmunds, M.G., 1992,
        \mnras, 255, 325
\reference P\'erez, E., 1996, \mnras, 290, 465
\reference Schaerer, D., 1996, \apj, 467, L17
\reference Schaerer, D., 1998, in ``Wolf-Rayet Phenomena in Massive Stars and Starburst 
Galaxies'', IAU Symp. 193, ASP Conf. Series, in press (astro-ph/9812357)
\reference Schaerer, D., Contini, T., Kunth D., 1999a, \aap, 341, 399 (SCK99)
\reference Schaerer, D., Contini, T., Pindao M., 1999b, \aaps, 136, 35
\reference Schaerer, D., Stasi\'nska, G., 1999, \aap, 345, L17 (SS99)
\reference Schaerer, D., Vacca, W. D. 1998, \apj, 497, 618 (SV98)
\reference Stasi\'nska, G., Schaerer, D., 1999, \aap, submitted
\reference Stevens, I.R., Strickland, D.K., 1998, \mnras, 301, 215
\reference Vacca, W.D., Conti, P.S., 1992, \apj,401, 543
\reference Vanbeveren, D., et al., 1998, NewA 3, 443
\reference Walsh, J.R., Roy, J-R., 1987, \apj, 319, L57
\reference Williams, P.M., 1995, IAU Symp. 163, 335

\end{references}
\end{document}